\documentclass[12pt,preprint]{aastex}

\shorttitle{Abundances in M107}
\shortauthors{O'Connell et al.}

\begin{document}

\title{Chemical Abundances of Red Giant Stars in the Globular Cluster M107 (NGC 
6171)}

\author{
Julia  E. O'Connell\altaffilmark{1},
Christian I. Johnson\altaffilmark{2,3,4},
Catherine A. Pilachowski\altaffilmark{3}, and
Geoffrey Burks\altaffilmark{1,5}
}

\altaffiltext{1}{Department of Mathematics and Physics, College of Arts and Sciences, Tennessee State University, Boswell Science Hall, Nashville, TN 37209-1561, USA; joconnell@mytsu.tnstate.edu, burks@coe.tsuniv.edu}

\altaffiltext{2}{UCLA Division of Astronomy, 475 Portola  Plaza, Physics and Astronomy Building 3-548, Los Angeles, CA 90095, USA; cijohnson@astro.ucla.edu}

\altaffiltext{3}{National Science
Foundation Astronomy and Astrophysics Postdoctoral Fellow}

\altaffiltext{4}{Department of Astronomy, Indiana University,
Swain West 319, 727 East Third Street, Bloomington, IN 47405--7105, USA; catyp@astro.indiana.edu}

\altaffiltext{5}{Center of Excellence in Information Systems, Tennessee
State University, 3500 John Merritt Blvd, Box 9501, Research and Sponsored Programs 242, 
Nashville, Tennessee 37209-1561, USA; burks@coe.tsuniv.edu}

\begin{abstract}

We present chemical abundances of Al and several Fe--Peak and neutron--capture elements for 13 red giant branch stars in the Galactic globular cluster NGC 6171 (M107). The abundances were determined using equivalent width and spectrum synthesis analyses of moderate resolution (R $\sim$15,000), moderate signal--to--noise ratio ($\langle$S/N$\rangle$$\sim$80) spectra obtained with the WIYN telescope and Hydra multifiber spectrograph. A comparison between photometric and spectroscopic effective temperature estimates seems to indicate a reddening value of E(B--V)=0.46 may be more appropriate for this cluster than the more commonly used value of E(B--V)=0.33. Similarly, we found that a distance modulus of (m--M)$_{\rm V}$$\approx$13.7 provided reasonable surface gravity estimates for the stars in our sample. Our spectroscopic analysis finds M107 to be moderately metal--poor with $\langle$[Fe/H]$\rangle$=--0.93 and also exhibits a small star--to--star metallicity dispersion ($\sigma$=0.04). These results are consistent with previous photometric and spectroscopic studies. Aluminum appears to be moderately enhanced in all program stars ($\langle$[Al/Fe]$\rangle$=+0.39, $\sigma$=0.11). The relatively small star--to--star scatter in [Al/Fe]  differs from the trend found in more metal--poor globular clusters, and is more similar to what is found in clusters with [Fe/H]$\ga$--1. The cluster also appears to be moderately r-- process enriched with $\langle$[Eu/La]$\rangle$=+0.32 ($\sigma$ = 0.17). 

\end{abstract}


\section{INTRODUCTION}
The old paradigm that globular clusters represent single, coeval stellar
populations has been overturned by the discovery of multiple, discrete
populations existing in seemingly ``normal\rq\rq{} clusters (e.g., see Renzini
2008; Piotto 2009; Milone et al. 2010 for recent reviews).  While it has
long been known that essentially all globular clusters exhibit significant
star--to--star abundance variations for the elements carbon to aluminum
(e.g., see Gratton et al. 2004 and references therein), the connection
between the light element variations and existence of multiple populations
is only now becoming more clear.  It is now thought that most clusters
contain (at least) two separate generations of stars.  These populations
exhibit identical [Fe/H]\footnote{We adopt the standard notations
[A/B]$\equiv$log(N$_{\rm A}$/N$_{\rm B}$)$_{\rm star}$--log(N$_{\rm
A}$/N$_{\rm B}$)$_{\sun}$ and log $\epsilon$(A)$\equiv$log(N$_{\rm
A}$/N$_{\rm H}$)+12.0 for elements A and B.} ratios, but differ in their
light element abundances (e.g., Carretta et al. 2009a).  The first
generation or ``primordial\rq\rq{} stars reflect the light element abundance
patterns produced by type II supernovae (SNe), which are nearly identical
to the metal--poor halo composition.  The second generation stars appear
to have formed from gas that experienced varying degrees of
high--temperature proton--capture nucleosynthesis, and are therefore in
general O/Mg--poor and Na/Al--rich compared to the typical halo field
star.  Interestingly, the second generation tends to be the dominant
population in most clusters, and the number of first generation stars
retained is likely a function of cluster mass (Carretta et al. 2009a).
While the more massive clusters tend to exhibit the most extreme light
element abundance and multiple population characteristics, smaller
clusters like M107, which do not appear to contain a significant fraction
of extremely O/Mg--poor and Na/Al--rich stars, may be useful probes for
determining the processes which produce the second generation in globular
clusters.

The Galactic globular cluster M107 is of relatively average mass
($\sim$10$^{\rm 5}$ M$_{\rm \sun}$; Piatek et al. 1994), but is a factor of
two more metal--rich than the average globular cluster (Harris 1996;
updated 2010\footnote{The catalog can be accessed at
http://physwww.physics.mcmaster.ca/$\sim$harris/mwgc.dat.}).  A compilation of
multiple photometric and moderate resolution spectroscopic analyses
(Pilachowski et al. 1981; Smith \& Perkins 1982; Smith \& Manduca 1983;
Zinn \& West 1984; Carretta \& Gratton 1997; Ferraro et al. 1999; Carretta
et al. 2009a, 2009b) yields a metallicity value between [Fe/H]=--0.83 and --1.07.
However, most of the spectroscopic measurements are based on small sample
sizes ($\la$ 5 stars).  Although globular clusters tend to exhibit a wide
range in horizontal branch (HB) morphologies at a given metallicity, the
HB of M107 is dominated by red HB and RR Lyrae stars (Sandage \& Katem
1964; Dickens \& Rolland 1972; Sandage \& Roques 1982; Da Costa et al.
1984; Ferraro et al. 1991; Cudworth et al. 1992) which is clearly reflected
in the (B--R)/(B+V+R) HB ratio estimate by Lee et al. (1994) of
--0.76$\pm$0.08.  M107 lacks a significant population of blue HB and blue
hook stars that are typically found in some of the more massive clusters
exhibiting the largest light element abundance variations and may suggest
that this cluster did not experience strong helium enrichment.

While the [Fe/H], [O/Fe], and [Na/Fe] ratios have been determined for
$\sim$30 red giant branch (RGB) stars in M107, chemical abundances for Al
and heavier $\alpha$ elements are only available for $\sim$5 stars, and
the neutron--capture element abundances have never been explored.
Therefore, we present for the first time moderate resolution spectroscopic
abundances of Al, Ti, Sc, Ni, Fe, La, and Eu for 13 RGB stars in this
cluster.  In section 2 we describe the selection of stars for observation
and data reduction.  Section 3 contains the radial velocity measurements
and cluster membership evaluations for individual stars.  Section 4
describes the procedures for estimating model atmosphere parameters and
measuring the chemical abundances.  Lastly, in section 5 we outline and
discuss the results and provide a summary in section 6.

\section{OBSERVATIONS AND REDUCTIONS}

The observations for all cluster giants were taken at Kitt Peak National Observatory on May 14, 2000 using the WIYN 3.5m telescope instrumented with the Hydra multi--fiber positioner and bench spectrograph. All spectra were obtained with a single Hydra configuration that employed the 2$\arcsec$ red fibers, 316 line mm$^{\rm -1}$ echelle grating and red camera, achieving a resolving power of R($\lambda$/$\Delta$$\lambda$)$\approx$15,000. The spectrograph setup was centered near 6660~\AA, and the full wavelength coverage spanned from $\sim$6460--6860~\AA. Target stars were selected based on photometry from Sandage \& Katem (1964), with colors suggesting their location on or near the RGB. The coordinates used in generating the Hydra configuration were taken from the USNO Image and Catalogue Archive\footnote{The Catalogue Archive Service can be found at http://www.nofs.navy.mil/data/fchpix/}. The final sample includes 13 RGB stars spanning a V magnitude range of 13.23--14.66, which corresponds to a luminosity range from the RGB tip down to approximately 1 magnitude above the level of the  HB (V$_{\rm HB}$ $\approx$15.7; Buonnano et al. 1989; Ferraro et al. 1991; see also Figure 1). 

Basic data reductions were carried out using the standard IRAF\footnote{IRAF is distributed by the National Optical Astronomy Observatories, which are operated by the Association of Universities for Research in Astronomy, Inc., under cooperative agreement with the National Science Foundation.} routines. Specifically, \emph{ccdproc} was used to apply the bias level correction and trim the overscan region. The IRAF task \emph{dohydra}
was employed for aperture tracing, scattered light and cosmic ray removal,
extraction of the one--dimensional spectra, flat--fielding, wavelength
calibration (based on a ThAr comparison source), and sky subtraction. The extracted spectra were then co--added to increase the S/N of the final spectra and continuum fit using a low order spline function.  The S/N of the combined spectra ranged from $\sim$60--110.

\section{RADIAL VELOCITY MEASUREMENTS AND CLUSTER MEMBERSHIP}

Cluster membership was confirmed by comparing radial velocity measurements to the mean value of  --34.23 km s$^{-1}$ found by Pryor \& Meylan (1993).  All radial velocities for this study were determined via the IRAF task \emph{fxcor}, and corrected for the Earth\rq{}s motion using \emph{rvcorrect}.  A proper motion study by Cudworth et al. (1992) presents membership probabilities for all stars selected for analysis, with the exception of star 201. Table 1 provides radial velocity measurements and associated uncertainties for each program star, as well as Cudworth\rq{}s membership probabilities. The average radial velocity of --31.8 km s$^{\rm -1}$ and small velocity dispersion ($\sigma$=2.4 km s$^{\rm -1}$ found here are in agreement with previous studies (e.g., Pryor \& Meylan 1993; Piatek et al. 1994)

Note that Smith \& Hesser (1986) exclude star F as a cluster member based on DDO photometry and identified it as a possible foreground dwarf.  However, Cudworth et al. (1992) assigned the star a high membership probability (98$\%$), and we find star F to have a radial velocity that is reasonably consistent with the cluster average at --37.1 km s$^{\rm -1}$. Although the radial velocity of star F is $\sim$2$\sigma$ outside the cluster mean, it has an effective temperature, surface gravity, and metallicity that are all consistent with the star being a \emph{bona fide} member. Therefore, we have included it in our analysis. 

\section{ANALYSIS}

We have analyzed a small sample of RGB stars in M107 for elemental abundances of Al, neutron--capture, and Fe--peak elements in the range of Al to Eu II. IRAF\rq{}s \emph{splot} package was used to measure equivalent widths (EWs) with a single-line EW analyses for unblended lines, and a blended--line function for heavily blended lines or lines subject to hyperfine splitting. The wavelength range of observed spectra is from $\sim$6460-6860 {\AA}. Effective temperatures and surface gravities for individual stars were initially estimated using  the cluster\rq{}s distance modulus and (V--K)$_{\rm 0}$ color indices obtained from photometric data. Although Sandage \& Katem (1964) provide photometry for all target stars, photometry for initial T$_{\rm eff}$ estimates was taken from the more recent proper motion study by Cudworth et al. (1992). Star 201 was not included in this study, but Dickens \& Rolland (1972) provide colors for 201 transformed from Sandage \& Katem (1964).  An iterative LTE stellar line analysis program was used to further modify T$_{\rm eff}$ and microturbulence (v$_{\rm t}$) via spectroscopic analyses. Table 2 shows the results of an assessment with respect to abundance sensitivity and associated uncertainties in adopted model atmosphere parameters for all elements considered in this study. 

\subsection{Model Stellar Atmospheres}

Initial T$_{\rm eff}$ estimates for individual stars were determined through use of the empirical V--K$_{\rm S}$ color--temperature relation described by Alonso et al. (1999, and erratum from 2001). The V--band photometry was obtained from Cudworth et al. (1992) and Dickens \& Rolland (1972), and the K$_{\rm S}$--band data were taken from the Two Micron All Sky Survey (2MASS) database (Skrutskie et al. 2006). A color excess value of E(B--V)=0.33 (Webbink 1985; Harris 1996), which is in agreement with the Cudworth et al. (1992) estimate, was initially adopted in order to correct for interstellar reddening and extinction. However, we found that applying this reddening correction produced T$_{\rm eff}$ values that were at least 150--200K lower than the T$_{\rm eff}$ estimates derived spectroscopically by imposing excitation equilibrium (see Figure 2). Further investigation of this problem revealed that the color excess for NGC 6171 is not particularly well constrained, with literature values ranging from E(B--V)$\approx$0.25--0.50 (e.g., Smith \& Hesser 1986; Salaris \& Weiss 1997). Dutra \& Bica (2000) noticed a similar inconsistency between their derived value of E(B--V)=0.45, based on 100$\micron$ dust emission, and previously published estimates. 

While our data do not permit an explicit measure of interstellar reddening along the clusters line--of--sight, we do find that a near 1:1 correlation between photometric and and spectroscopic T$_{\rm eff}$ estimates for cluster giants can be achieved if one assumes a reddening near the upper limit of E(B--V)$\approx$0.46 (see Figure 2). Since this larger E(B--V) value is also found in the Schlegel et al. (1998) dust maps, which were accessed via the NED Coordinate Transformation \& Galactic Extinction Calculator\footnote{http://nedwww.ipac.caltech.edu/forms/calculator.html}, we used an average value of E(B--V)=0.46 in the final T$_{\rm eff}$ calculations. Furthermore, use of the online extinction calculator permitted a rough examination into the prospect of differential reddening across our observed field, which could be an issue given the cluster\rq{}s low Galactic latitude (\emph{b}=23$\degr$). Fortunately, the star--to--star reddening variation did not exceed 0.02 mag, and therefore no additional corrections were applied.

Surface gravities were calculated using the standard relation,  \begin{equation}
log(g) = 0.40(M_{bol.}-M_{bol.\sun}+log(g_\sun)+4(log(T/T_\sun))+log(M/M_\sun),
\end{equation} and assumed a stellar mass of of 0.8 M$_{\sun}$.  Stellar atmospheres were modeled without convective overshoot by interpolating in the ATLAS9 grid\footnote{Kurucz model atmospheres can be found at http://kurucz.harvard.edu/grids.html} (Castelli et al. 1997). The absolute bolometric magnitudes (M$_{\rm bol.}$) were determined by applying the V--band
bolometric corrections from Alonso et al. (1999; their equations 17 and
18) to the absolute V--band magnitudes estimated from the distance modulus (m--M)$_{\rm V}$=13.76 (Shetrone et al. 2009).  In a similar fashion to
the reddening estimate, a wide range of distance modulus estimates for
this cluster appear in the literature and span from (m--M)$_{\rm V}$=15.06
(e.g., Harris 1996) to (m--M)$_{\rm V}$=13.76 (Shetrone et al. 2009).
However, we chose the smallest available distance modulus because the
larger distance moduli yielded surface gravity values that appeared too
low for each star's metallicity and position on the color--magnitude
diagram.

Initial model atmospheres were calculated with a metallicity of
[Fe/H]$\approx$--1, which is consistent with previous estimates (e.g.,
Smith \& Manduca 1983; Pilachowski 1984; Zinn \& West 1984; Ferraro et al.
1991; Carretta et al. 2009a, 2009b), and also assumed a microturbulence value of 2
km s$^{\rm -1}$ for all stars.  These values were further refined through
an iterative process that primarily focused on finalizing the
microturbulence value by removing trends in Fe I abundance as a function
of reduced width [log(EW/$\lambda$)].  A summary of our final model
atmosphere parameters and photometric indicies is provided in Table 3.

\subsection{Equivalent Width Analyses, Hyperfine Structure, and Spectrum Synthesis}

All element abundances, with the exception of Al, were derived by
equivalent width (EW) measurements using IRAF\rq{}s \emph{splot} package.
Suitable lines were chosen both by visual inspection and comparison to the
Hinkle et al. (2000) Arcturus atlas, which combines a side--by--side
profile of the solar and Arcturus spectra.  Given the moderate resolution
of our spectra, we chose lines for analysis that were not expected to be
severely blended.  While the abundances of Ti, Fe, Ni, and La were determined
by employing the \emph{abfind} driver in the 2002 version of the LTE line
analysis code MOOG (Sneden 1973), the abundances of Al, Sc, and Eu
were either determined via the \emph{synth} spectrum synthesis driver (Al)
or the blended line \emph{blends} driver (Sc and Eu).

For Al, we chose to derive the abundances using full spectrum synthesis of
the 6690-6700 \AA\ window because both the 6696 and 6698 \AA\ Al lines are
moderately blended with nearby metal and CN lines.  For Sc, La, and Eu,
the abundance derivation requires taking into account hyperfine structure
and/or isotopic broadening.  While both Sc and La have only one long--lived,
stable isotope ($^{\rm 45}$Sc and $^{\rm 139}$La), Eu has two ($^{\rm 151}$Eu
and $^{\rm 153}$Eu) that are present in nearly equal proportions.  Therefore,
our input linelists for Sc and Eu made use of the the hyperfine/isotope data
from Prochaska \& McWilliam (2000) and Lawler et al. (2001), respectively.
Although no hyperfine linelist exists for the 6774 \AA\ La II line used
here, we applied the empirical correction given in Johnson \& Pilachowski
(2010; equation A1) to our measured EWs.  The final EWs and abundance
ratios, cited as relative to Fe I, are provided in Tables 4 and 5,
respectively.

\section{RESULTS AND DISCUSSION}

\subsection{Al Abundances}

Large star--to--star light element abundance variations are a ubiquitous
feature of globular clusters (e.g., see reviews by Kraft 1994; Gratton et
al. 2004).  While it is understood that these abundance patterns, in
particular those involving the elements between carbon and aluminum, are
the result of proton--capture nuclear reactions, the exact production
site(s) is (are) not well established.  Evolved red giants have deep
convective envelopes that can mix proton--capture cycled material from a
star's interior to its photosphere, and this mechanism is clearly
responsible for the first dredge--up phenomenon (e.g., Iben 1965).
However, observations of similar abundance variations involving heavier
elements, from O to Al, in globular cluster stars near the main--sequence
and turn off (e.g., Cannon et al. 1998; Gratton et al. 2001; Cohen et al.
2002; Briley et al. 2004a, 2004b; Boesgaard et al. 2005) suggest pollution
must play a key role as well.  The most commonly suggested pollution sites
tend to be either rapidly rotating, massive stars (e.g., Maeder \& Meynet
2006) or $\sim$5--8 M$_{\rm \sun}$ AGB stars (e.g., Ventura \& D'Antona
2009).  While the AGB scenario tends to be the most commonly accepted, it
is likely that both massive and intermediate mass stars play key roles in
determining the light element composition of globular cluster stars (see
also Renzini 2008 for a recent review).  Since Al is the heaviest element
that generally exhibits a large abundance range in globular clusters, it
requires the highest temperatures to be produced ($\ga$5$\times$10$^{\rm
7}$ K) in significant quantities.  These temperatures are not expected to
be reached at the bottom of the convective envelope in low mass stars with
[Fe/H]$\approx$-1, and therefore Al can be used as a tracer for the amount
of pollution experienced by M107 stars.

We find the individual [Al/Fe] ratios to be enhanced by an average of
$+$0.39 dex with a relatively small dispersion of $\sigma$=0.11 dex.
While the full range of [Al/Fe] spans from $+$0.24 to +0.63 dex, only two
stars (J and 205) have [Al/Fe]$>$+0.5.  The enhancement of Al in star 205
is shown in Figure 3 where we overplot stars N and 205, which have
similar T$_{\rm eff}$, log(g), and [Fe/H], but differ in their [Al/Fe]
ratios by $\sim$0.3 dex.

In Figure 4 we show a box plot of the [Al/Fe] ratios for 13
globular clusters ranging in [Fe/H] from approximately --2.35 to --0.80.
While it is clear from Figure 4 that the overwhelming majority of
globular cluster stars have [Al/Fe]$>$0, there appears to be a significant
change in the [Al/Fe] abundance spreads for the more metal--rich clusters,
including M107.  The metal--poor, and generally more massive, clusters
tend to exhibit a full range of [Al/Fe] abundances spanning nearly a
factor of 10, but the clusters with [Fe/H]$\ga$--1.2 tend to exhibit
abundance spreads of only 0.1--0.5 dex.  This observation is not entirely
surprising, especially when considered in the context of the commonly
assumed paradigm that the light element abundance dispersions in globular
clusters are driven primarily by pollution from intermediate mass AGB
stars because theoretical Type II SNe and AGB yields tend to converge at
[Fe/H]$\ga$--1.2 (e.g., see Figure 22 in Johnson \& Pilachowksi 2010 and
references therein).  This means a cluster like M107, forming from gas
polluted by Type II SNe and AGB stars with metallicities near
[Fe/H]$\sim$--1, should not exhibit the same large [Al/Fe] spread seen in 
stars forming from gas polluted by more metal--poor progenitors.
Therefore, the observed small [Al/Fe] variations observed in M107 are
consistent with its metallicity.  However, the average [Al/Fe]=$+$0.39 is
at least 0.3 dex lower than the predicted yields of the $\sim$5--6.5
M$_{\rm \sun}$ AGB stars that are commonly assumed to be the primary
polluters in globular clusters (e.g., Ventura \& D'Antona 2009; but see
also Karakas 2010).  The moderate Al enhancement in M107 suggests that the
gas from which these stars formed did not exceed an AGB/Type II SN
pollution ratio of roughly 20$\%$/80$\%$, respectively.  This result is
compatible with the observed modest extension of M107's O--Na
anticorrelation seen in Carretta et al. (2009a).

\subsection{$\alpha$, Fe--Peak, and Neutron--Capture Elements}

Although Ti is often enhanced in globular clusters like the lighter, true
$\alpha$ elements (e.g., Mg and Ca), its exact nucleosynthetic origin is
unclear.  However, M107 does not appear to be an exception as both the
[Ti I/Fe] and [Ti II/Fe] ratios indicate that cluster stars are enhanced
by an average [Ti/Fe]=$+$0.40 with a relatively small star--to--star
dispersion ($\sigma$=0.10).  Similarly, the Fe--peak elements, traced here
by Sc and Ni, are typically not enhanced in globular cluster stars and
tend to exhibit small star--to--star dispersions.  We find that M107 fits
this trend as Ni exhibits no enhancements on average with
$\langle$[Ni/Fe]$\rangle$=0.00 ($\sigma$=0.09) and Sc also appears only
moderately enhanced at $\langle$[Sc/Fe]$\rangle$=+0.13 with a small
star--to--star dispersion ($\sigma$=0.09).  The enhancement of [Ti/Fe] and
solar--scaled abundance ratios of [Sc/Fe] and [Ni/Fe] are clearly
illustrated in Figure 5, where we show a box plot of all elements
measured in this study.

Most stable isotopes of elements heavier than the Fe--peak are produced
through either the rapid (r) or slow (s) neutron--capture process (e.g.,
see review by Sneden et al. 2008).  In general, the heavier elements
synthesized via the main component of the s--process (e.g., Ba and La) are
believed to be primarily produced in lower mass ($\sim$1--3 M$_{\rm
\sun}$) thermally pulsing AGB stars over timescales $\ga$5$\times$10$^{\rm
8}$ yrs.  In contrast, the exact origin of the r--process is unknown, but
it is believed to be associated with core collapse SNe and therefore
enrichment should occur on a rapid timescale of $\la$5$\times$10$^{\rm 7}$
yrs (e.g., see review by Truran et al. 2002).  R--process production is
often traced through the element Eu, which is produced almost exclusively
by the r-process.

While the star--to--star dispersion for neutron--capture elements in
globular clusters is typically larger than that observed for the $\alpha$
and Fe--peak elements (e.g., see Roederer 2011 and references therein), it
is almost always smaller than the variations observed for the lighter
elements C through Al.
However, on average most globular clusters have [Eu/La]$\ga$$+$0.2 (e.g.,
Gratton et al. 2004), which suggests that the clusters formed rapidly and
before a significant amount of s--process enrichment could occur.  M107
exhibits this same trend with
$\langle$[La/Fe]$\rangle$=$+$0.41 ($\sigma$=0.12),
$\langle$[Eu/Fe]$\rangle$=$+$0.73 ($\sigma$=0.13), and
$\langle$[Eu/La]$\rangle$=$+$0.32 ($\sigma$=0.17).  Although the [Eu/Fe]
ratio exhibits the largest abundance range out of all the elements
included in this study, the [Eu/Fe] interquartile range is not appreciably
different than the other elements.  This suggests that the cluster formed
from gas that was well mixed and exhibited a nearly homogeneous
composition.  Lastly, the negligible s--process signature indicates that
low and intermediate mass AGB stars did not contribute strongly
to the cluster's primordial composition, which further supports the
observed relatively small light element abundance variations observed here
and in previous studies.   

\section{SUMMARY}

We present for the first time moderate resolution spectroscopic abundances of Fe, Al, Ti, Sc, Ni, La, and Eu for 13 RGB stars in the globular cluster NGC 6171 (M107).  All data for this study were obtained at Kitt Peak National Observatory with the  WIYN 3.5m telescope and Hydra multifiber spectrograph using a moderate resolution (R$\sim$15,000) echelle grating. The coadded spectra have a $\langle$S/N$\rangle$ $\sim$80 and cover a wavelength range from $\sim$6460-6860 \AA.  Program stars range in luminosity from the RGB tip to $\sim$1 magnitude above the level of the HB. 

Effective temperatures and surface gravities for individual stars were estimated using the cluster\rq{}s distance modulus and (V--K)$_{\rm 0}$ color indices obtained from photometric data. An iterative LTE stellar line analysis code was employed to further modify T$_{\rm eff}$ and microturbulence (v$_{\rm t}$) via spectroscopic analyses. With the exception of Al, abundances were determined by equivalent width (EW) analyses. For Al we chose to derive abundances using spectrum synthesis to eliminate possible contamination from nearby CN and metal lines. Input linelists were used for Sc and Eu to provide hyperfine structure and/or isotope broadening corrections. An empirical correction was applied to our La II EW measurements as no hyperfine linelist exists for this line.  

Given the low galactic latitude of this cluster and close relative proximity to the galactic center (\emph{b}=23$\degr$ and R$_{\rm GC}$=3.3 kpc, respectively), interstellar reddening and extinction can be a possible concern. Reddening values and distance moduli found in literature were not very well constrained, but by assuming a color excess value close to the upper limit found in literature, E(B--V) $\sim$0.46, we find a near 1:1 correlation between photometric and spectroscopic T$_{\rm eff}$ estimates. Similarly, we chose the smallest available distance modulus, (m--M)$_{\rm V}$=13.76, because the larger distance moduli yielded surface gravity values that appeared too low for each star\rq{}s metallicity and position on the color magnitude diagram. 

We confirm that M107 is moderately metal--rich, with average [Fe/H]=--0.93 ($\sigma$=0.04), which is consistent with previous photometric and spectroscopic studies. Program stars indicate a small star--to--star metallicity spread of 0.12 dex suggesting M107 is a \emph{bona fide} monometallic cluster. Carretta et al. (2009a) finds a similar star--to--star spread in [Fe/H],  0.18 dex, and the same $\sigma$ value, 0.04, for 33 stars in this cluster. The HB of M107 is dominated by red HB stars ((B--R)/(B+V+R)=--0.76$\pm$0.08) and RR Lyrae variables, which would not be unexpected for its metallicity, and lacks a significant population of blue HB and blue hook stars. This may indicate that M107 did not experience strong helium enrichment typically demonstrated by some of the more massive clusters that also tend to exhibit the largest light element abundance variations.

We find that the [Al/Fe] ratio is enhanced in all cluster stars at
$\langle$[Al/Fe]$\rangle$=$+$0.39 ($\sigma$=0.11) with only two stars
having [Al/Fe]$>$+0.5.  The ``baseline\rq\rq{} [Al/Fe]=$+$0.24 is consistent
with predicted yields from Type II SNe, but the average [Al/Fe]
enhancements are well below the  theoretical yields from similar
metallicity, intermediate mass AGB stars.  The small star--to--star
[Al/Fe] variations observed in M107 follow the trend observed for other
clusters of similar metallicity.  Similarly, we find that M107 exhibits
``typical\rq\rq{} globular cluster abundance ratios with respect to the heavier
elements.  The surrogate $\alpha$ element tracer Ti is enhanced with
$\langle$[Ti/Fe]$\rangle$=$+$0.40 ($\sigma$=0.10), and the two Fe--peak
elements Sc and Ni exhibit nearly solar--scaled abundance ratios with
$\langle$[Sc/Fe]$\rangle$=$+$0.13 ($\sigma$=0.09) and
$\langle$[Ni/Fe]$\rangle$=0.00 ($\sigma$=0.09).  Finally,
the neutron--capture elements indicate that M107 is r--process rich
($\langle$[La/Fe]$\rangle$=$+$0.41 ($\sigma$=0.12),
$\langle$[Eu/Fe]$\rangle$=$+$0.73 ($\sigma$=0.13), and
$\langle$[Eu/La]$\rangle$=$+$0.32 ($\sigma$=0.17)) and therefore likely
formed quite rapidly.  The relatively small star--to--star element
variations in this cluster suggest it did not experience a significant
amount of self--enrichment.

\acknowledgements

We extend gratitude to Diane Harmer for obtaining all observations used in this study. This publication makes use of data products from the Two Micron All Sky Survey, which is a joint project of the University of Massachusetts and the Infrared Processing and Analysis Center/California Institute of Technology, funded by the National Aeronautics and Space Administration and the National Science Foundation.This research has also made use of the NASA/IPAC Extragalactic Database (NED) which is operated by the Jet Propulsion Laboratory, California Institute of Technology, under contract with the National Aeronautics and Space Administration. This work was supported in part by the National Science Foundation through an REU Site Program grant to Tennessee State University, AST--0453557, and Indiana University, AST--0453437. JEO is also grateful for the continuation of support for this study from the Tennessee State University Center of Excellence for Research and Sponsored Programs through the National Science Foundation grant AST--0958267. Support of the College of Arts and Sciences at Indiana University Bloomington for CIJ is gratefully acknowledged. This material is based upon work supported by the National Science Foundation under award No. AST--1003201 to C.I.J.

\clearpage

\begin{deluxetable}{ccccc}
\tablenum{1}
\tablecolumns{5}
\tablewidth{0pt}

\tablecaption{Radial Velocity and Membership Information}
\tablehead{
\colhead{Star\tablenotemark{a}} &
\colhead{V$_{\rm R}$}   &
\colhead{Error} &
\colhead{$\sigma$ from Mean}  &
\colhead{Mem. Prob.\tablenotemark{b}}   \\
\colhead{}      &
\colhead{(km s$^{\rm -1}$)}      &
\colhead{(km s$^{\rm -1}$)}      &
\colhead{(km s$^{\rm -1}$)}      &
\colhead{}
}

\startdata
F	&	$-$37.1	&	1.1	&	2.1	&	98	\\
G	&	$-$32.0	&	0.9	&	1.5	&	97	\\
H	&	$-$29.7	&	1.3	&	3.1	&	97	\\
J	&	$-$30.9	&	0.7	&	2.3	&	94	\\
K	&	$-$29.9	&	0.8 	&	3.0	&	98	\\
L	&	$-$33.2	&	0.9	&	0.6	&	98	\\
N	&	$-$34.3	&	0.8	&	0.2	&	98	\\
O	&	$-$34.3	&	1.5	&	0.2	&	94	\\
R	&	$-$29.7	&	1.4	&	3.6	&	97	\\
201	&	$-$31.0	&	0.9	&	2.2	&	\nodata	\\
205	&	$-$32.1	&	0.9	&	1.4	&	96	\\
273	&	$-$30.1	&	1.5	&	2.9	&	89	\\
278	&	$-$29.8	&          0.7	&	3.1	&	98	\\

\hline
\multicolumn{5}{c}{Cluster Mean Values}    \\
\hline
Average	&	$-$31.8	&	1.1	&	\nodata	&	\nodata	\\
Median	&	$-$31.0	&	0.9	&	\nodata	&	\nodata	\\
Std. Dev.	&	2.4	&	0.3	&	\nodata	&	\nodata	\\

\enddata

\tablenotetext{a}{Star identifiers are from Sandage \& Katem (1964).}
\tablenotetext{b}{Membership probabilities are from Cudworth et al. (1992).}

\end{deluxetable}

\clearpage

\begin{deluxetable}{ccccc}
\tablenum{2}
\tablecolumns{5}
\tablewidth{0pt}

\tablecaption{Abundance Sensitivity to Model Atmosphere Parameters}
\tablehead{
\colhead{Element} &
\colhead{$\Delta$T$_{\rm eff}$ $\pm$ 100}   &
\colhead{$\Delta$log $g$ $\pm$ 0.30} &
\colhead{$\Delta$[M/H] $\pm$ 0.30}  &
\colhead{$\Delta$v$_{t}$ $\pm$ 0.30}   \\
\colhead{}      &
\colhead{(K)}      &
\colhead{(cgs)}      &
\colhead{(dex)}      &
\colhead{(dex)}
}

\startdata
Fe I	&	$\pm$0.09	&	$\pm$0.04	&	$\pm$0.02	&	$\pm$0.10	\\
Al I	&	$\pm$0.07	&	$\pm$0.01	&	$\pm$0.01	&	$\pm$0.02	\\
Ti I	&	$\pm$0.17	&	$\pm$0.00	&	$\pm$0.04	&	$\pm$0.06	\\
Ti II	&	$\pm$0.03	&	$\pm$0.11	&	$\pm$0.09	&	$\pm$0.02	\\
Sc II	&	$\pm$0.04	&	$\pm$0.17 	&	$\pm$0.15	&	$\pm$0.04	\\
Ni I	&	$\pm$0.03	&	$\pm$0.04	&	$\pm$0.04	&	$\pm$0.06	\\
La II	&	$\pm$0.03	&	$\pm$0.10	&	$\pm$0.11	&	$\pm$0.15	\\
Eu II	&	$\pm$0.02	&	$\pm$0.09	&	$\pm$0.10	&	$\pm$0.01	\\

\enddata

\end{deluxetable}

\clearpage

\begin{deluxetable}{ccccccccccc}
\tablenum{3}
\tablecolumns{11}
\tablewidth{0pt}

\tablecaption{Photometry and Model Atmosphere Parameters}
\tablehead{
\colhead{Star\tablenotemark{a}}	&
\colhead{V\tablenotemark{b}}       &     
\colhead{B--V}	    &
\colhead{J}	    &
\colhead{H}	    &
\colhead{K$_{\rm S}$}	    &
\colhead{T$_{\rm eff}$}	&
\colhead{log $g$}	&
\colhead{[Fe/H]}	&
\colhead{v$_{\rm t}$}	&
\colhead{S/N}	\\
\colhead{}	&
\colhead{}      &
\colhead{}      &
\colhead{}      &
\colhead{}      &
\colhead{}      &
\colhead{(K)}      &
\colhead{(cgs)}      &
\colhead{}      &
\colhead{(km s$^{\rm -1}$)}      &
\colhead{}
}

\startdata
F	&	13.39	&	1.70	&	9.995	    &	9.118       &	8.923       &	4090	&	0.90	&	$-$0.96	&	2.15	&	75	\\
G	&	13.50	&	1.66	&	10.191	    &	9.352       &	9.111       &	4150	&	0.90	&	$-$0.93	&	1.70	&	80	\\
H	&	13.84	&	1.61	&	10.589	    &	9.742       &	9.536       &	4200	&	1.05	&	$-$0.96	&	1.95	&	60	\\
J	&	13.97	&	1.58	&	10.909	    &	10.121     &	9.902	    &	4360	&	1.25	&	$-$0.92	&	1.70	&	110	\\
K	&	14.04	&	1.48	&	11.049	    &	10.282	    &	10.108	    &	4450	&	1.45	&	$-$0.95	&	2.10	&	80	\\
L	&	14.04	&	1.47	&	11.020	    &	10.252	    &	10.071	    &	4450	&	1.50	&	$-$0.87	&	1.70	&	85	\\
N	&	14.26	&	1.53	&	11.219	    &	10.398	    &	10.256	    &	4420	&	1.45	&	$-$0.86	&	1.75	&	90	\\
O	&	14.36	&	1.44	&	11.424	    &	10.653	    &	10.473	    &	4490	&	1.65	&	$-$0.91	&	1.90	&	75	\\
R	&	14.66	&	1.28	&	11.963	    &	11.301	    &	11.096	    &	4780	&	2.10	&	$-$0.96	&	1.95	&	70	\\
201	&	14.44	&	1.27	&	11.731	    &	11.110	    &	10.870	    &	4790	&	1.85	&	$-$0.98	&	1.85	&	70	\\
205	&	14.56	&	1.45	&	11.598	    &	10.821	    &	10.656	    &	4485	&	1.60	&	$-$0.93	&	1.90	&	75	\\
273	&	13.23	&	1.81	&	9.605	    &	8.703	    &	8.438	    &	3950	&	0.70	&	$-$0.97	&	1.80	&	70	\\
278	&	14.14	&	1.48	&	11.110	    &	10.338	    &	10.105	    &	4400	&	1.45	&	$-$0.95	&	1.80	&	75	\\
\enddata

\tablenotetext{a}{Star identifiers are from Sandage \& Katem (1964).}
\tablenotetext{b}{Photometry for all stars except 201 is from Cudworth et al. (1992).\\ Photometry for star 201 is from Dickens \& Rolland (1972).}

\end{deluxetable}

\clearpage

\begin{deluxetable}{ccccccccccccccccc}
\tablenum{4}
\tablecolumns{17}
\tablewidth{0pt}

\tabletypesize{\scriptsize}
\rotate
\tablecaption{Equivalent Widths\tablenotemark{a,b}}
\tablehead{
\colhead{Wavelength}	&
\colhead{Species}      &
\colhead{E.P.}      &
\colhead{log gf}      &
\colhead{F\tablenotemark{c}}      &
\colhead{G}      &
\colhead{H}      &
\colhead{J}      &
\colhead{K}      &
\colhead{L}      &
\colhead{N}      &
\colhead{O}      &
\colhead{R}      &
\colhead{201}      &
\colhead{205}      &
\colhead{273}      &
\colhead{278}      \\
\colhead{\AA}	&
\colhead{}	&
\colhead{eV}      &
\colhead{}      &
\colhead{}      &
\colhead{}      &
\colhead{}      &
\colhead{}      &
\colhead{}      &
\colhead{}      &
\colhead{}      &
\colhead{}      &
\colhead{}      &
\colhead{}      &
\colhead{}      &
\colhead{}      &
\colhead{}
}

\startdata
6696.03	&	Al I 	&	3.14	&	$-$1.57	&	synth	&	synth	&	synth	&	synth	&	synth	&	synth	&	synth	&	synth	&	synth	&	synth	&	synth	&	synth	&	synth	\\
6698.66	&	Al I	&	3.14	&	$-$1.89	&	synth	&	synth	&	synth	&	synth	&	synth	&	synth	&	synth	&	synth	&	synth	&	synth	&	synth	&	synth	&	synth	\\
6604.60	&	Sc II	&	1.36	&	$-$1.48	&	91	&	91	&	84	&	84	&	72	&	68	&	88	&	78	&	61	&	58	&	79	&	94	&	83	\\
6554.23	&	Ti I	&	1.44	&	$-$1.16	&	132	&	135	&	\nodata	&	99	&	85	&	80	&	88	&	85	&	67	&	\nodata	&	100	&	142	&	94	\\
6556.07	&	Ti I	&	1.46	&	$-$1.10	&	142	&	137	&	124	&	103	&	80	&	95	&	97	&	82	&	52	&	52	&	119	&	159	&	\nodata	\\
6743.12	&	Ti I	&	0.90	&	$-$1.65	&	\nodata	&	160	&	\nodata	&	\nodata	&	\nodata	&	\nodata	&	\nodata	&	105	&	73	&	61	&	\nodata	&	183	&	118	\\
6559.57	&	Ti II	&	2.05	&	$-$2.30	&	73	&	\nodata	&	75	&	71	&	68	&	68	&	73	&	60	&	68	&	\nodata	&	\nodata	&	63	&	\nodata	\\
6606.97	&	Ti II	&	2.06	&	$-$2.79	&	39	&	48	&	42	&	\nodata	&	37	&	37	&	\nodata	&	36	&	40	&	34	&	48	&	41	&	44	\\
6475.63	&	Fe I	&	2.56	&	$-$3.01	&	\nodata	&	111	&	\nodata	&	\nodata	&	\nodata	&	\nodata	&	\nodata	&	\nodata	&	\nodata	&	\nodata	&	92	&	\nodata	&	\nodata	\\
6481.87	&	Fe I	&	2.28	&	$-$3.08	&	152	&	\nodata	&	139	&	115	&	125	&	113	&	121	&	120	&	\nodata	&	80	&	109	&	141	&	114	\\
6494.99	&	Fe I 	&	2.40	&	$-$1.24	&	288	&	254	&	269	&	230	&	\nodata	&	236	&	236	&	230	&	192	&	191	&	241	&	265	&	233	\\
6498.95	&	Fe I	&	0.96	&	$-$4.69	&	180	&	154	&	158	&	136	&	138	&	\nodata	&	\nodata	&	131	&	88	&	81	&	\nodata	&	\nodata	&	132	\\
6574.25	&	Fe I	&	0.99	&	$-$5.02	&	155	&	132	&	\nodata	&	107	&	112	&	\nodata	&	110	&	\nodata	&	63	&	63	&	\nodata	&	132	&	108	\\
6592.92	&	Fe I	&	2.73	&	$-$1.47	&	219	&	189	&	196	&	178	&	189	&	173	&	180	&	172	&	151	&	147	&	\nodata	&	194	&	178	\\
6593.88	&	Fe I	&	2.43	&	$-$2.42	&	181	&	156	&	162	&	145	&	152	&	141	&	144	&	142	&	112	&	109	&	\nodata	&	171	&	143	\\
6597.57	&	Fe I	&	4.79	&	$-$0.95	&	51	&	\nodata	&	48	&	\nodata	&	42	&	46	&	\nodata	&	45	&	31	&	\nodata	&	42	&	49	&	\nodata	\\
6608.04	&	Fe I	&	2.28	&	$-$3.96	&	91	&	82	&	\nodata	&	66	&	\nodata	&	\nodata	&	61	&	51	&	28	&	28	&	53	&	90	&	61	\\
6609.12	&	Fe I	&	2.56	&	$-$2.69	&	148	&	130	&	134	&	115	&	\nodata	&	114	&	\nodata	&	114	&	93	&	90	&	109	&	138	&	115	\\
6646.96	&	Fe I	&	2.61	&	$-$3.96	&	54	&	50	&	47	&	\nodata	&	29	&	34	&	40	&	\nodata	&	\nodata	&	14	&	31	&	50	&	32	\\
6677.99	&	Fe I	&	2.69	&	$-$1.35	&	233	&	\nodata	&	215	&	188	&	203	&	188	&	197	&	192	&	162	&	150	&	193	&	209	&	192	\\
6703.57	&	Fe I	&	2.76	&	$-$3.01	&	106	&	\nodata	&	96	&	82	&	87	&	81	&	82	&	80	&	48	&	45	&	81	&	\nodata	&	\nodata	\\
6710.32	&	Fe I	&	1.48	&	$-$4.83	&	\nodata	&	94	&	99	&	81	&	75	&	    \nodata 	&	77	&	66	&	38	&	32	&	72	&	112	&	74	\\
6750.16	&	Fe I	&	2.42	&	$-$2.62	&	169	&	145	&	\nodata	&	132	&	138	&	130	&	130	&	127	&	108	&	97	&	125	&	161	&	135	\\
6806.85	&	Fe I	&	2.73	&	$-$3.10	&	104	&	92	&	94	&	\nodata	&	80	&	78	&	78	&	74	&	44	&	\nodata	&	71	&	96	&	\nodata	\\
6482.80	&	Ni I	&	1.93	&	$-$2.79	&	118	&	\nodata	&	\nodata	&	\nodata	&	100	&	88	&	\nodata	&	80	&	75	&	\nodata	&	80	&	111	&	95	\\
6532.88	&	Ni I	&	1.93	&	$-$3.47	&	\nodata	&	69	&	\nodata	&	\nodata	&	\nodata	&	\nodata	&	\nodata	&	\nodata	&	\nodata	&	\nodata	&	\nodata	&	\nodata	&	\nodata	\\
6586.31	&	Ni I	&	1.95	&	$-$2.81	&	\nodata	&	108	&	104	&	97	&	96	&	86	&	92	&	80	&	74	&	64	&	\nodata	&	119	&	90	\\
6643.63	&	Ni I	&	1.68	&	$-$2.01	&	205	&	180	&	177	&	168	&	174	&	156	&	168	&	153	&	\nodata	&	123	&	149	&	191	&	160	\\
6767.78	&	Ni I	&	1.83	&	$-$2.17	&	173	&	153	&	152	&	137	&	159	&	139	&	145	&	132	&	\nodata	&	110	&	124	&	\nodata	&	139	\\
6772.32	&	Ni I	&	3.66	&	$-$0.96	&	67	&	\nodata	&	69	&	70	&	\nodata	&	66	&	\nodata	&	\nodata	&	50	&	45	&	47	&	\nodata	&	\nodata	\\
6774.33	&	La II	&	0.13	&	$-$1.75	&	59	&	64	&	63	&	30	&	38	&	50	&	41	&	46	&	25	&	21	&	33	&	74	&	51	\\
6645.12	&	Eu II	&	1.37	&	$+$0.20	&	50	&	69	&	51	&	59	&	45	&	60	&	56	&	52	&	39	&	27	&	64	&	67	&	48	\\
\enddata

\tablenotetext{a}{The designation ``Synth" indicates a synthetic spectrum
comparison method was used.}
\tablenotetext{b}{Equivalent widths are given in units of m\AA.}
\tablenotetext{c}{Star identifiers are from Sandage \& Katem (1964).}

\end{deluxetable}

\clearpage
\pagestyle{empty}
\setlength{\voffset}{0.25in}

\begin{deluxetable}{cccccccccccccccccccccc}
\tablenum{5}
\tablecolumns{22}
\tablewidth{0pt}

\tabletypesize{\scriptsize}
\rotate
\setlength{\tabcolsep}{0.06in}
\tablecaption{Measured Abundances}
\tablehead{
\colhead{Star\tablenotemark{a}}	&
\colhead{[Fe/H]}      &
\colhead{$\sigma$}      &
\colhead{N}      &
\colhead{[Al/Fe]}      &
\colhead{$\sigma$}      &
\colhead{N}      &
\colhead{[ScII/Fe]}      &
\colhead{$\sigma$}      &
\colhead{N}      &
\colhead{[Ti/Fe]$_{\rm avg.}$}      &
\colhead{$\sigma$}      &
\colhead{N}      &
\colhead{[Ni/Fe]}      &
\colhead{$\sigma$}      &
\colhead{N}      &
\colhead{[LaII/Fe]}      &
\colhead{$\sigma$}      &
\colhead{N}      &
\colhead{[EuII/Fe]}      &
\colhead{$\sigma$}      &
\colhead{N} 
}

\startdata
F	&	$-$0.96	&	0.02	&	14	&	$+$0.31	&	0.07	&	2	&	$+$0.01	&	\nodata	&	1	&	$+$0.27	&	0.04	&	4	&	$-$0.04	&	0.07	&	4	&	$+$0.25	&	\nodata	&	1	&	$+$0.49	&	\nodata	&	1	\\
G	&	$-$0.93	&	0.03	&	12	&	$+$0.36	&	\nodata	&	1	&	$+$0.09	&	\nodata	&	1	&	$+$0.49	&	0.01	&	4	&	$+$0.07	&	0.04	&	4	&	$+$0.40	&	\nodata	&	1	&	$+$0.76	&	\nodata	&	1	\\
H	&	$-$0.96	&	0.03	&	12	&	$+$0.40	&	0.12	&	2	&	$+$0.09	&	\nodata	&	1	&	$+$0.36	&	0.01	&	3	&	$-$0.08	&	0.05	&	4	&	$+$0.50	&	\nodata	&	1	&	$+$0.57	&	\nodata	&	1	\\
J	&	$-$0.92	&	0.04	&	12	&	$+$0.55	&	\nodata	&	1	&	$+$0.19	&	\nodata	&	1	&	$+$0.42	&	0.01	&	3	&	$+$0.06	&	0.08	&	4	&	$+$0.13	&	\nodata	&	1	&	$+$0.80	&	\nodata	&	1	\\
K	&	$-$0.95	&	0.03	&	12	&	$+$0.30	&	\nodata	&	1	&	$+$0.08	&	\nodata	&	1	&	$+$0.29	&	0.04	&	4	&	$+$0.06	&	0.05	&	4	&	$+$0.42	&	\nodata	&	1	&	$+$0.69	&	\nodata	&	1	\\
L	&	$-$0.87	&	0.02	&	11	&	$+$0.42	&	\nodata	&	1	&	$-$0.06	&	\nodata	&	1	&	$+$0.33	&	0.08	&	4	&	$-$0.02	&	0.04	&	5	&	$+$0.51	&	\nodata	&	1	&	$+$0.80	&	\nodata	&	1	\\
N	&	$-$0.86	&	0.05	&	12	&	$+$0.31	&	\nodata	&	1	&	$+$0.26	&	\nodata	&	1	&	$+$0.37	&	0.04	&	3	&	$+$0.05	&	0.05	&	3	&	$+$0.37	&	\nodata	&	1	&	$+$0.77	&	\nodata	&	1	\\
O	&	$-$0.91	&	0.06	&	13	&	$+$0.46	&	\nodata	&	1	&	$+$0.13	&	\nodata	&	1	&	$+$0.34	&	0.05	&	5	&	$-$0.11	&	0.02	&	4	&	$+$0.52	&	\nodata	&	1	&	$+$0.78	&	\nodata	&	1	\\
R	&	$-$0.96	&	0.05	&	13	&	$+$0.39	&	\nodata	&	1	&	$+$0.24	&	\nodata	&	1	&	$+$0.55	&	0.07	&	5	&	$+$0.13	&	0.08	&	3	&	$+$0.55	&	\nodata	&	1	&	$+$0.89	&	\nodata	&	1	\\
201	&	$-$0.98	&	0.05	&	13	&	$+$0.28	&	\nodata	&	1	&	$+$0.18	&	\nodata	&	1	&	$+$0.42	&	0.02	&	3	&	$+$0.00	&	0.07	&	4	&	$+$0.43	&	\nodata	&	1	&	$+$0.65	&	\nodata	&	1	\\
205	&	$-$0.93	&	0.06	&	12	&	$+$0.63	&	\nodata	&	1	&	$+$0.21	&	\nodata	&	1	&	$+$0.61	&	0.08	&	3	&	$-$0.19	&	0.05	&	4	&	$+$0.32	&	\nodata	&	1	&	$+$0.97	&	\nodata	&	1	\\
273	&	$-$0.97	&	0.07	&	13	&	$+$0.24	&	0.04	&	2	&	$+$0.09	&	\nodata	&	1	&	$+$0.35	&	0.09	&	5	&	$+$0.08	&	0.09	&	3	&	$+$0.38	&	\nodata	&	1	&	$+$0.67	&	\nodata	&	1	\\
278	&	$-$0.95	&	0.04	&	12	&	$+$0.40	&	\nodata	&	1	&	$+$0.21	&	\nodata	&	1	&	$+$0.46	&	0.06	&	3	&	$+$0.01	&	0.06	&	4	&	$+$0.55	&	\nodata	&	1	&	$+$0.69	&	\nodata	&	1	\\
\hline
\multicolumn{22}{c}{Cluster Mean Values}    \\
\hline
Average	&	$-$0.93	&	\nodata	&	\nodata	&	$+$0.39	&	\nodata	&	\nodata	&	$+$0.13	&	\nodata	&	\nodata	&	$+$0.40	&	\nodata	&	\nodata	&	$+$0.00	&	\nodata	&	\nodata	&	$+$0.41	&	\nodata	&	\nodata	&	$+$0.73	&	\nodata	&	\nodata	\\
Median	&	$-$0.95	&	\nodata	&	\nodata	&	$+$0.39	&	\nodata	&	\nodata	&	$+$0.13	&	\nodata	&	\nodata	&	$+$0.37	&	\nodata	&	\nodata	&	$+$0.01	&	\nodata	&	\nodata	&	$+$0.42	&	\nodata	&	\nodata	&	$+$0.76	&	\nodata	&	\nodata	\\
Std. Dev.	&	0.04	&	\nodata	&	\nodata	&	0.11	&	\nodata	&	\nodata	&	0.09	&	\nodata	&	\nodata	&	0.10	&	\nodata	&	\nodata	&	0.09	&	\nodata	&	\nodata	&	0.12	&	\nodata	&	\nodata	&	0.13	&	\nodata	&	\nodata	\\
\enddata

\tablenotetext{a}{Star identifiers are from Sandage \& Katem (1964).}

\end{deluxetable}

\newpage
\begin{figure}
\caption{Color--magnitude diagram of M107 with photometry from Cudworth et al. (1992) represented by both red and black filled circles. Program stars are indicated, with no overlap, by larger filled red circles. Photometry from Ferraro et al. (1991) plotted as blue diamonds.}
\epsscale{1.00}
\plotone{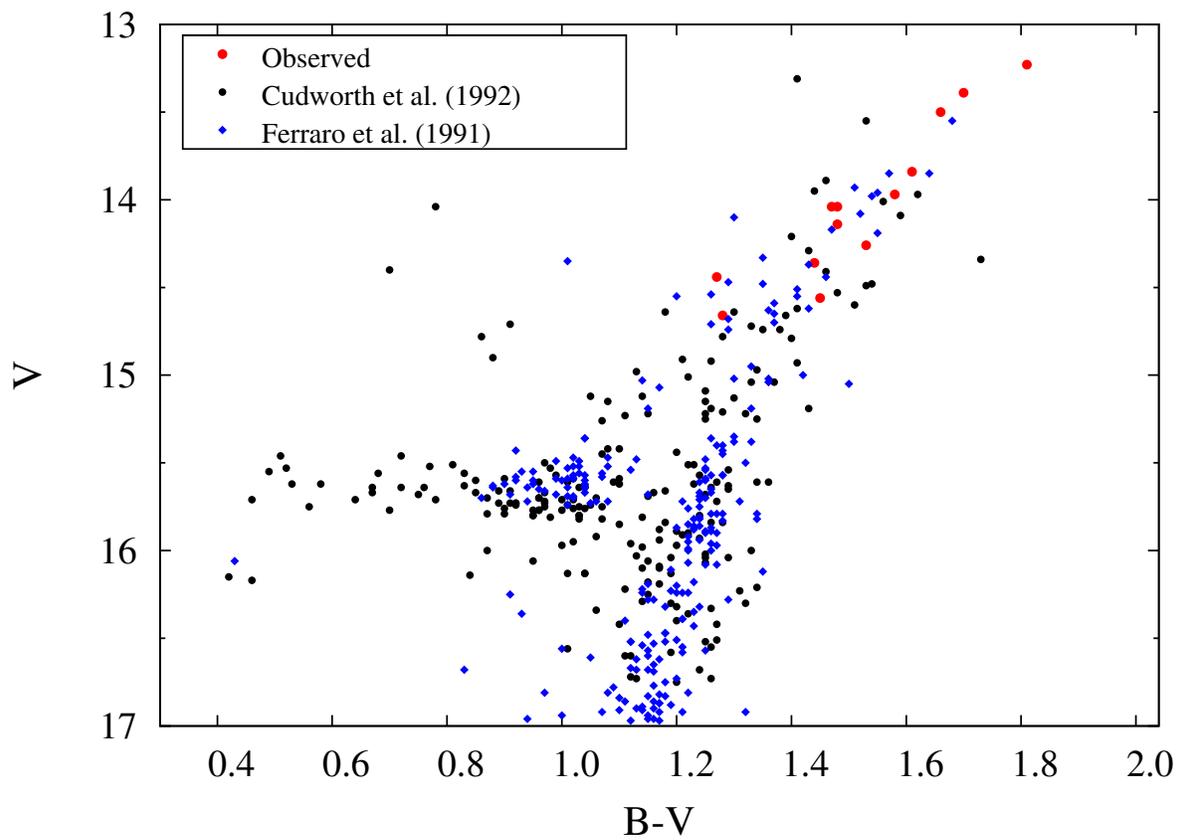}
\label{f1}
Note--  Photometry for star 201 is from Dickens \& Rolland (1972)
\end{figure}

\newpage
\begin{figure}
\caption{The dashed diagonal represents perfect agreement between photometric estimates of effective temperature and T$_{\rm eff}$ derived spectroscopically. Open circles depict stellar T$_{\rm eff}$ models using a reddening value more commonly found in literature for this cluster, E(B--V)=0.33. Filled circles along the diagonal represent the results of photometric T$_{\rm eff}$ estimates using E(B--V)=0.46 and our final effective temperatures derived for each star.}
\epsscale{1.10}
\hspace{-1 in}
\plotone{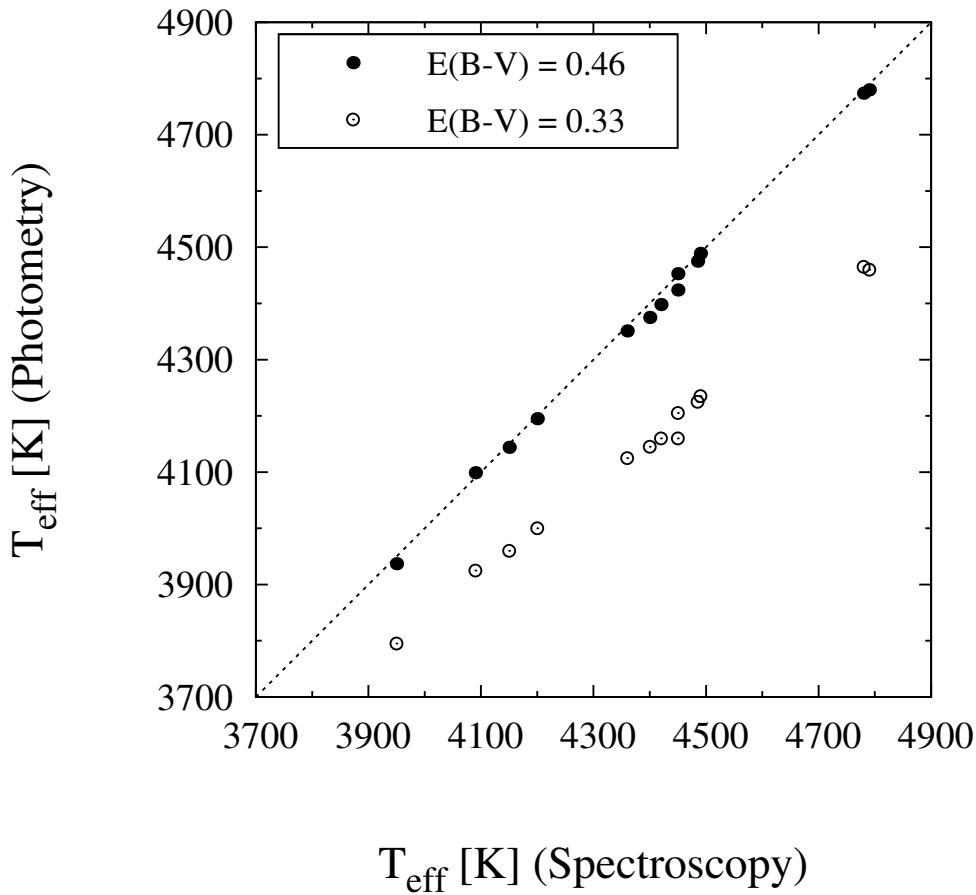}
\label{f2}
\end{figure}

\newpage
\begin{figure}
\caption{Spectral line profiles of stars 205 and N about Al doublet $\lambda\lambda$6696 \& 6698 \AA\  illustrating the star--to--star Al abundance dispersion. The two stars have similar stellar model atmospheres, yet marked differences in log $\epsilon$(Al) values.}
\epsscale{1.00}
\hspace{-0.5 in}
\plotone{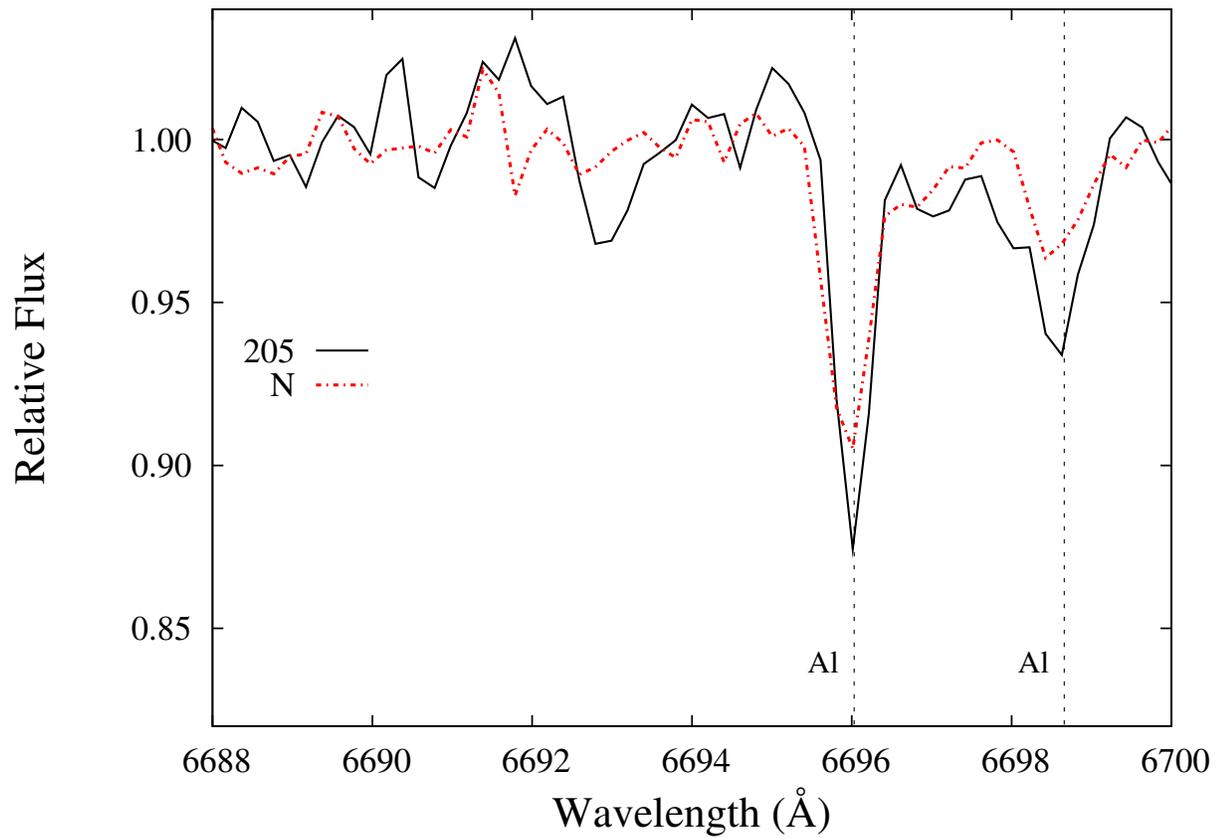}
\label{f3}
\end{figure}

\newpage
\begin{figure}
\caption{Box plot illustrating [Al/Fe] distribution in 13 galactic globular clusters --2.35 $\lesssim$ [Fe/H] $\lesssim$ --0.75. Clusters are plotted by increasing metallicity. The middle line of each box indicates the median abundance value, and the upper and lower box boundaries represent the third and first quartiles (75$^{\rm th}$ and 25$^{\rm th}$ percentile) of the data, respectively. The vertical lines represent the full range of abundance values. Suspected outliers (stars with abundances 1.5 times above the 3$^{\rm rd}$, or below the 1$^{\rm st}$ interquartile range) are designated by filled circles, and outliers (abundances 3.0 times above the 3$^{\rm rd}$, or below the 1$^{\rm st}$ interquartile range) are open circles. M80 from Cavallo et al. (2004); M13 from Johnson et al. (2005);  M30, M68, M55, M10, NGC 6752, M12, NGC 288, M4, M71 and NGC 104 from Carretta et al. (2009b); M107 from the current study.}
\epsscale{1.05}
\hspace{-0.75 in}
\plotone{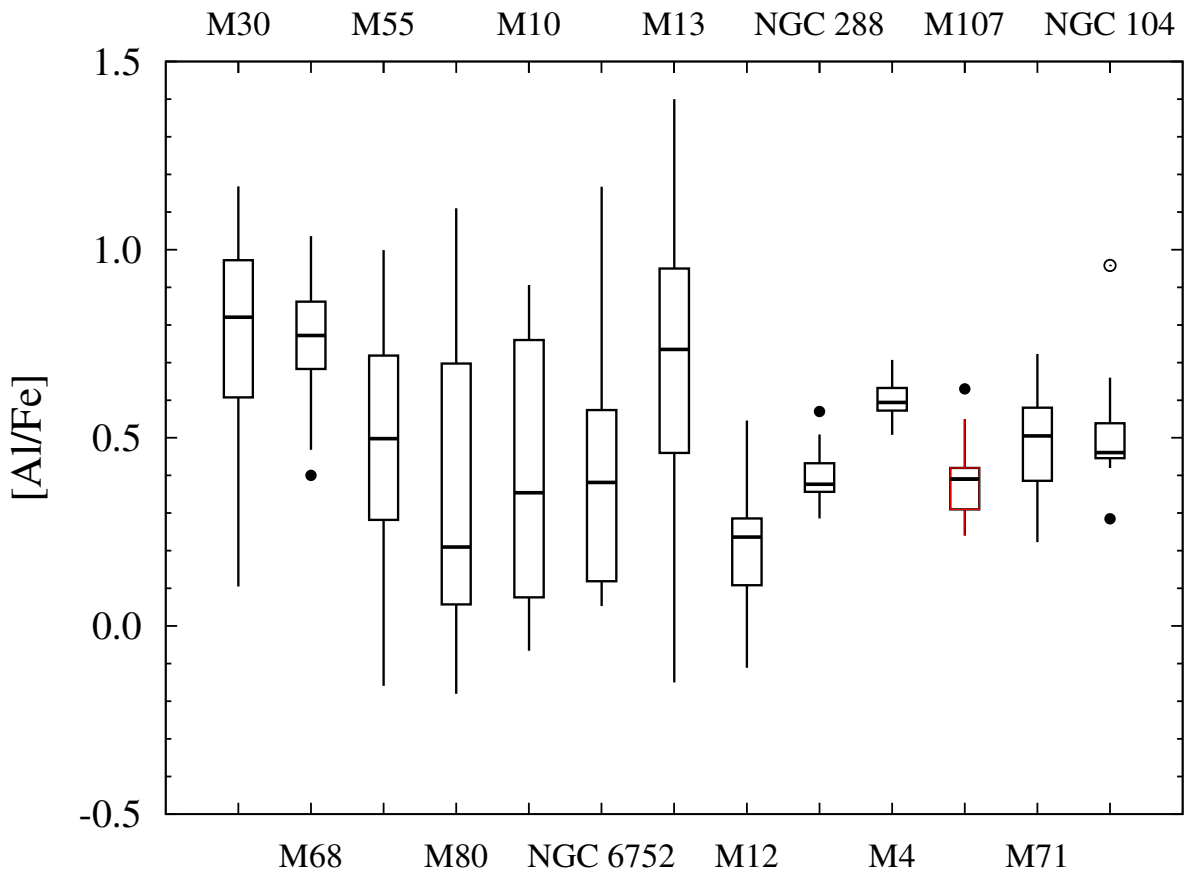}
\label{f4}
\end{figure}

\newpage
\begin{figure}
\caption{Box plot indicating the abundance distribution and star--to--star variation in the measured abundances relative to Fe I, with $\langle$Ti$\rangle$ as the mean value of Ti I and Ti II. The plotting designations are the same as Figure \ref{f4}.} 
\epsscale{1.00}
\hspace{-0.5 in}
\plotone{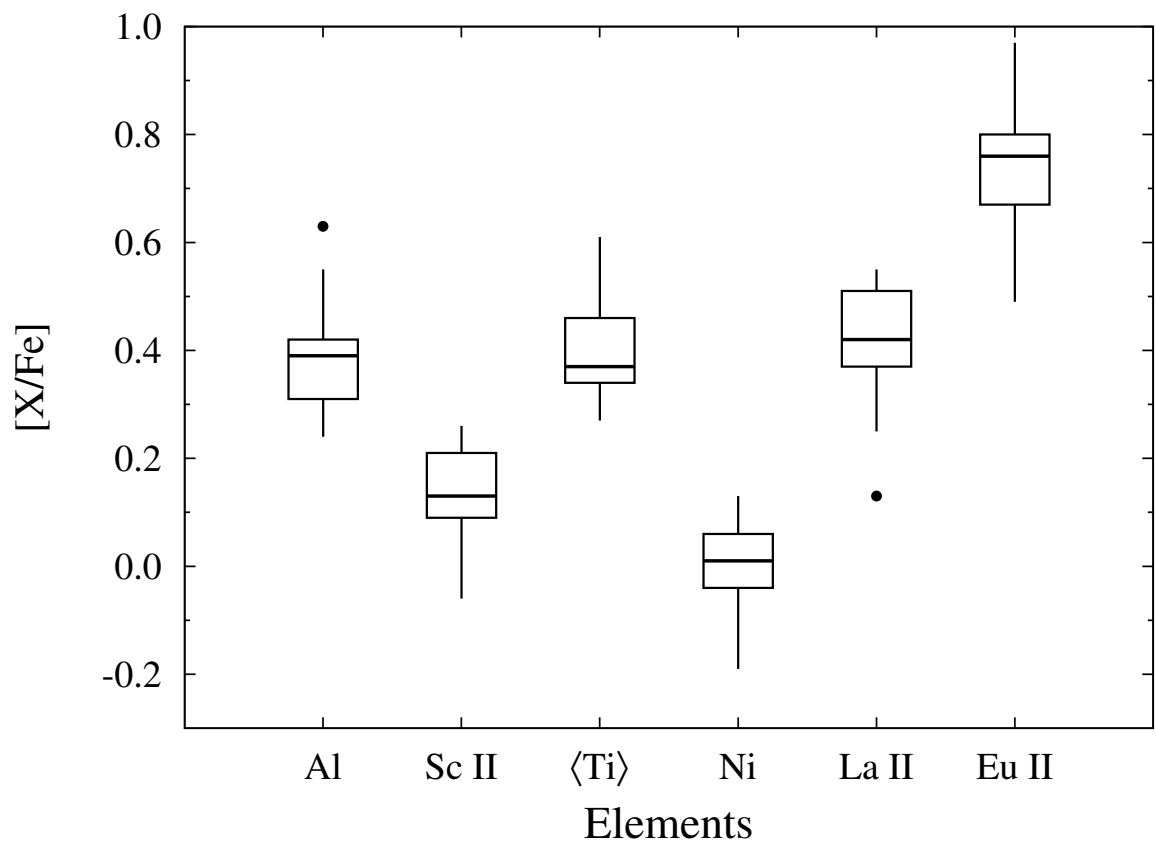}
\label{f5}
\end{figure}

\end{document}